\documentclass[superscriptaddress,reprint,amsmath,amssymb,aps]{revtex4-2}
\usepackage[english]{babel}
\usepackage{amsmath,amssymb,mathtools,float}
\usepackage{graphicx}
\usepackage{dcolumn}
\usepackage{bm}
\usepackage{xr}
\usepackage[version=4]{mhchem}
\usepackage{hyperref}
\usepackage[dvipsnames]{xcolor}
\usepackage{siunitx}
\DeclareSIUnit\angstrom{\text {Å}}
\DeclareSIUnit\molar{\text {M}}
\usepackage{layouts}
\usepackage{mathtools} 
\usepackage{amssymb}   
\usepackage{siunitx}   
\usepackage{soul}
\usepackage{float}

\usepackage{pdfpages}

\makeatletter
\AtBeginDocument{\let\LS@rot\@undefined}
\makeatother

\usepackage{tikz}

\usepackage{tabularx}   
\usepackage{longtable} 
\usepackage{xltabular} 
\usepackage{booktabs}   

\usepackage[capitalize]{cleveref}

\makeatletter

\begin{document}

\title{Convection can enhance the capacitive charging of porous
electrodes}

\author{Aaron D. Ratschow}
\thanks{A.D.R. and A.J.W. contributed equally to this work.}
\affiliation{
Technical University of Darmstadt,
Department of Mechanical Engineering, Institute for
Nano- and Microfluidics, Darmstadt D-64287, Germany
}
\affiliation{Department of Physics at Interfaces, Max Planck Institute
for Polymer Research, Mainz D-55128, Germany;}
 
\author{Alexander J. Wagner}
\thanks{A.D.R. and A.J.W. contributed equally to this work.}
\affiliation{
Technical University of Darmstadt,
Department of Mechanical Engineering, Institute for
Nano- and Microfluidics, Darmstadt D-64287, Germany
}
 
\author{Mathijs Janssen}
\email{mathijs.a.janssen@nmbu.no}
\affiliation{Institute of Physics, Faculty of Science and Technology,
Norwegian University of Life Sciences, \AA s 1433,
Norway}

\author{Steffen Hardt}
\email{hardt@nmf.tu-darmstadt.de}
\affiliation{
Technical University of Darmstadt,
Department of Mechanical Engineering, Institute for
Nano- and Microfluidics, Darmstadt D-64287, Germany
}

\date{\today}

\begin{abstract}

Charge transport in porous electrodes is foundational for modern energy storage technologies like supercapacitors, fuel cells, and batteries. Supercapacitors in particular rely solely on storing energy in charged pores. Here, we simulate the charging of a single electrolyte-filled pore using the modified Poisson-Nernst-Planck and Navier-Stokes equations. We find that electroconvection
can substantially speed up the charging dynamics. 
We uncover the fundamental mechanism of electroconvection during pore charging through an analytical model that predicts the induced flow field and the electric current arising due to convection. 
Our findings suggest that convection is especially important in the limit of slender pores with thin electric double layers, and becomes significant beyond a certain threshold voltage that is an inherent electrolyte property. 
\end{abstract}

\maketitle

To harness the full potential of renewable but intermittent sources like solar and wind, next generation batteries, supercapacitors, and other storage devices are needed \cite{Raza.2018,Karthikeyan.2021,Sahin.2022,Sharma.2023,Chanut.2023}.
Batteries and supercapacitors both use porous electrodes to store energy; batteries do so chemically, supercapacitors through the formation of electric double layers (EDLs). 
Porous electrode charging is both multiscale and multiphysics, and all theoretical models for it have focussed on particular scales or phenomena while simplifying or ignoring others \cite{Jeanmairet.2022,Wu.2022,Kondrat.2023}. 
Traditionally, models for porous electrode charging either focus on the device scale and treat pore charging in an averaged way \cite{Biesheuvel.2010, Newman.1962,Henrique.2024}, or focus on single pores. 
Initially, single pores were described through equivalent circuits such as the transmission line circuit \cite{Levie.1963,Levie.1964,Posey.1966}. 
Single pores were then connected in networks to predict whole-electrode charging, but such models, despite their popularity, do not give a first principles understanding of pore charging and rely on (several) unexplained fit parameters.  
In the last two decades, the ion transport mechanisms of diffusion and electromigration that underlie pore charging have been studied in Poisson-Nernst-Planck (PNP) simulations of simple pore geometries \cite{Sakaguchi.2007,Mirzadeh.2014,Gupta.2020,Henrique.2021,Henrique.2022,Aslyamov.2022,pedersen2023equivalent, Yang.2022} and idealized porous structures  \cite{Lian.2020,Henrique.2024}.
Still, such first-principles models predict the characteristic time with which porous electrodes charge within an order of magnitude at best \cite{Lian.2020}.
In this article, we show that electroconvection---solvent flow due to motion of ions in electric fields--- can substantially affect single pore charging.

Electroconvection is well established in electrodialysis, capacitive desalination, and desalination shock waves \cite{Alizadeh.2017,Alizadeh.2017b,Alizadeh.2019}. It was theoretically predicted \cite{Rubinstein.2000} and experimentally observed \cite{Rubinstein.2008} in electrodialysis, where up to \si{mm/s} velocities occur in reservoirs adjacent to membranes with fixed surface charges \cite{Nikonenko.2017}. It can even cause unstable or chaotic vortices \cite{Davidson.2014} and generally enhances cross-membrane charge transport in the overlimiting current regime \cite{Kim.2007,Yossifon.2008}. Bulk electroconvection influences electrodeposition of charged colloids \cite{Ammam.2012}. 
A recent work \cite{Ratschow.2022} showed that electroconvection can also arise in a conical nanopore between two reservoirs. 
When a voltage is applied between the pore wall and a counterelectrode in the reservoir, the resulting electric field acts on the emerging space charge in the pore, the diffuse part of the developing EDL. 
The mechanism unveiled in \cite{Ratschow.2022} differs from previous instances of electroconvection in pores in that the applied voltage caused both the space charge in the liquid and the electric field acting on that space charge. The same mechanism should, in principle, also affect the charging of pores in supercapacitors, which is what we set out to study here.

In this work, we report results of finite-element simulations of the fully coupled modified PNP \cite{Kilic.2007b} and Navier-Stokes (NS) equations. 
We find that for aqueous electrolytes, the influence of electroconvection on 
a pore's charging dynamics can become especially relevant at moderate and higher wall potentials compared to the
thermal voltage $k_\mathrm{B} T/e\approx \SI{25}{\milli\volt}$ ($e$: elementary charge, $k_\mathrm{B}$: Boltzmann's constant, 

\noindent $T=\SI{295}{K}$: temperature). At high potentials, the effect is strongest  for slender pores with thin EDLs.
We explain the fundamental mechanism of electroconvection during pore charging and demonstrate it through 
an analytical model that reproduces the simulated velocity field and convective current at low and moderate wall potentials by combining exact solutions and heuristic elements. 

\section*{Setup}
\begin{figure}[t]
    \centering
    \includegraphics[width=1\linewidth]{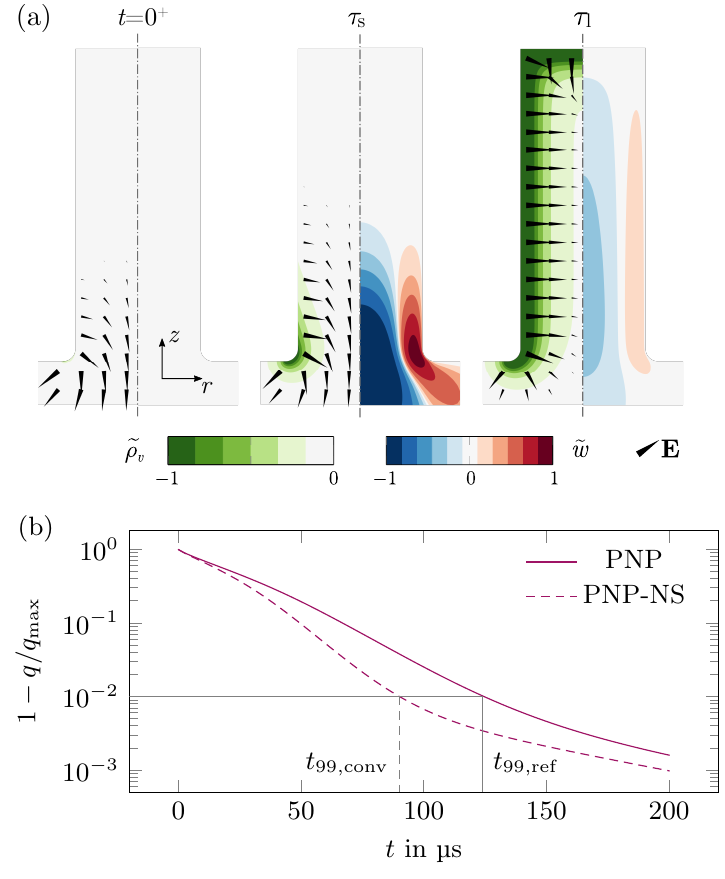}
    \caption{Velocity and charge concentration fields as well as charge relaxation dynamics inside a pore. (a) Simulation results for $L_\mathrm{p}/r_\mathrm{p}=r_\mathrm{p}/\lambda=5$ and $\Tilde{\zeta}=1$. Scaled charge density $\Tilde{\rho}_v$, electric field $\mathbf{E}$ (left split), and axial velocity $\Tilde{w}$ (right split)
    directly after switching on the wall potential and at the short $\tau_\mathrm{s}$  and long timescale $\tau_\mathrm{l}$.
    (b) Typical relaxation of the pore charge $q$ for $\Tilde{\zeta}=10$ as predicted by models accounting for diffusion and electromigration (PNP), and additionally for electroconvection (PNP-NS). 
    The charging times $t_{99,\mathrm{conv}}$ and $t_{99,\mathrm{ref}}$ are indicated.}\label{Fig1}
\end{figure}

We consider a two-dimensional axisymmetric computational domain representing a single cylindrical dead-end pore of radius $r_\mathrm{p}\in\{1,2,5,10,20,50\}\si{nm}$ 
and length $L_\mathrm{p}\in\{5,10,20,25,40,50,100,200,250,400,500,1000,2500\}\si{nm}$, 
connected to a reservoir, represented by a quadrant of radius $r_\infty$, large enough to suppress finite-size effects [Supporting Information (SI) 1.B., \cite{Haynes.2017,Richardson.1911,Probstein.1994}]. The edge at the pore entrance is blunted with a radius $r_\mathrm{blunt}=\SI{1}{nm}$ (\cref{Fig1}a). 

We solve for the electric potential $\phi$, ion concentrations $c_\pm$, velocity $\mathbf{u}$, and pressure $p$ using the modified PNP equations to account for ion crowding at higher potentials \cite{Kilic.2007b}, alongside the incompressible NS equations. 
Inertial forces and transient terms are included in the numerical code of the NS equations, but are negligible due to small Reynolds and Strouhal numbers (SI 3.B). The equations are coupled through the charge density $\rho_v=e\left(c_+-c_-\right)$, convective ion flux $\mathbf{u}c_\pm$, and the electric body force $\rho_v\mathbf{E}$ in the P, NP, and NS equations, respectively (Materials and Methods). 
Here, $\mathbf{E}=-\nabla\phi$ is the electric field. 
The predictive power of the modified PNP-NS equations for experiments has been confirmed in the literature \cite{Chen.2024}. We additionally demonstrate it by comparison with induced-charge electroosmosis experiments \cite{Feng.2017} (SI 1.D). 

The pore and reservoir are filled with a symmetric, monovalent, binary electrolyte solution of constant permittivity $\varepsilon=80\varepsilon_0$ ($\varepsilon_0$: vacuum permittivity), mass density $\rho_\mathrm{m}=\SI{1000}{kg/m^3}$, viscosity $\mu=\SI{1}{mPa\,s}$, ion volume $a^3=(\SI{0.3}{nm})^3$, ion diffusivity $D=\SI{1e-9}{m^2/s}$, and bulk concentration $c_0$, yielding a Debye length $\lambda=\sqrt{\varepsilon kT/(2e^2c_0})=\SI{1}{nm}$. 
The values are based on the properties of an aqueous \ce{KCl} solution \cite{Haynes.2017}. 
Far from the pore, the reservoir is grounded, at concentration $c_0$, and stress-free. The pore and reservoir walls are considered non-slipping and impermeable to ions and solvent. Hydrodynamic slip has been reported in nanochannels and would further enhance electroconvection 
\cite{CottinBizonne.2005,Bocquet.2010,Secchi.2016}.
Initially, the pore walls, considered to have no native zeta potential, are grounded. At time $t=0$, a step potential at the pore walls is switched on, according to $\zeta(t) = \Theta(t)\zeta_0$ [$\Theta(\cdot)$: Heaviside function].

\section*{Influence of electroconvection}
To assess the influence of electroconvection on pore charging, we compare numerical solutions of the fully coupled modified PNP-NS equations (index `$\mathrm{conv}$') and of the modified PNP equations without fluid flow, $\mathbf{u}=\mathbf{0}$ (index `$\mathrm{ref}$'). We determine the pore charge $q$ by integrating the surface charge density 
over the interior pore surface $A_\mathrm{p}$, excluding the blunting, $q=\int_{A_\mathrm{p}}\mathbf{n}\cdot\varepsilon\mathbf{E}\mathrm{d}A$. Following \cite{Mirzadeh.2014}, we then evaluate the time $t_{99}=t(q=0.99q_\mathrm{max})$ at which the pore reaches $99\%$ of its maximum charge $q_\mathrm{max}=q(t\to\infty)$, visualized in \cref{Fig1}b. The relative deviation in charging time $\Delta \tilde{t}=(t_{99,\mathrm{ref}}-t_{99,\mathrm{conv}})/t_{99,\mathrm{conv}}$ measures the influence of electroconvection. 
More specifically, $\Delta \tilde{t}$ quantifies the overestimation of charging times that is caused by neglecting electroconvection in pore charging models. 

\begin{figure}[t]
    \centering
    \includegraphics[width=1\linewidth]{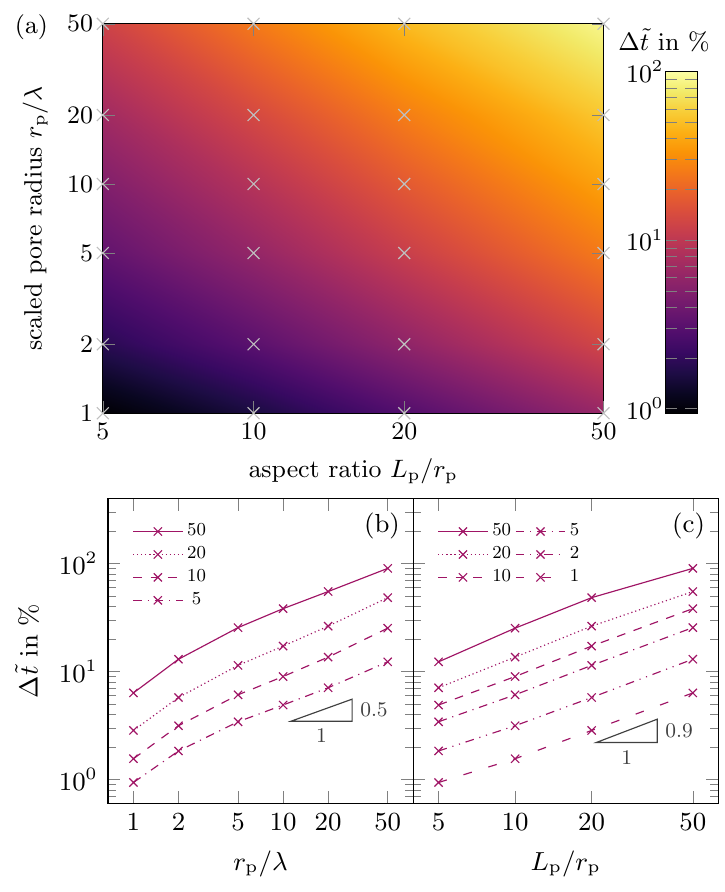}  
    \caption{Simulation results in the strongly nonlinear regime $\Tilde{\zeta}=10$. (a) Heatmap of the relative deviation in charging time $\Delta\Tilde{t}$ over the scaled pore radius $r_\mathrm{p}/\lambda$ and aspect ratio $L_\mathrm{p}/r_\mathrm{p}$. Symbols mark calculated data points with linear interpolation in between. (b) $\Delta\Tilde{t}$ along vertical cut lines through the heatmap at different horizontal positions with indication of scaling exponents. (c) Analogous to (b), but using horizontal cut lines along different vertical positions.}
	\label{Fig2}
\end{figure}

The heatmap in \cref{Fig2}a shows $\Delta \Tilde{t}$ in the plane of the scaled pore radius $r_\mathrm{p}/\lambda$ and aspect ratio $L_\mathrm{p}/r_\mathrm{p}$ for a wall potential $\Tilde{\zeta}=\zeta_0 e/(k_\mathrm{B} T)=10$. 
Here, the modified PNP equations are strongly nonlinear, leading to salt uptake and ion crowding near the pore's surface \cite{Kilic.2007b,Biesheuvel.2010}. In this strongly nonlinear regime, $\Delta\Tilde{t}$ increases with $r_\mathrm{p}/\lambda$ and $L_\mathrm{p}/r_\mathrm{p}$ over the entire considered parameter space. 
Electroconvection affects slender pores with thin EDLs most, with $\Delta\Tilde{t}$ exceeding $90\%$.
\Cref{Fig2}b-c show that the increase in $\Delta\Tilde{t}$ over the scaled pore radius and aspect ratio approximately follows a power law with exponents of $0.5$ and $0.9$, respectively. 
Presumably, $\Delta\Tilde{t}$ surpasses $100\%$ at higher scaled pore radii and aspect ratios.

\begin{figure}[t]
    \centering
    \includegraphics[width=1\linewidth]{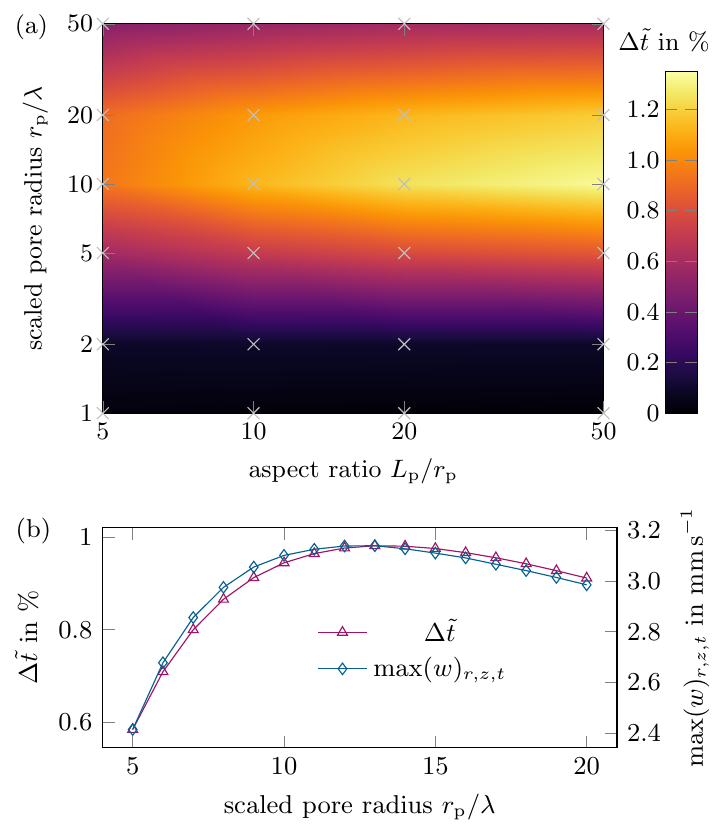}
    \caption{Simulation results in the weakly nonlinear regime $\Tilde{\zeta}=1$. (a) Heatmap of the relative deviation in charging time $\Delta\Tilde{t}$ over the scaled pore radius $r_\mathrm{p}/\lambda$ and aspect ratio $L_\mathrm{p}/r_\mathrm{p}$. Symbols mark calculated data points with linear interpolation in between. The influence of convection is largely independent of $L_\mathrm{p}/r_\mathrm{p}$ but shows a maximum at $r_\mathrm{p}/\lambda\approx10$.
    (b) $\Delta\Tilde{t}$ over the scaled pore radius $r_\mathrm{p}/\lambda$, with additional datapoints compared to (a), and maximum induced velocity $\mathrm{max}(w)_{r,z,t}$, for $L_\mathrm{p}/r_\mathrm{p}=5$. 
    }
    \label{Fig3}
\end{figure}

For smaller applied wall potentials, electroconvection affects pore charging  less, see \cref{Fig3}a where $\Tilde{\zeta}=1$ (and SI 2, Fig. S6 where $\Tilde{\zeta}=4$). 
The heatmap in \cref{Fig3}a, showing the influence of geometry,  differs substantially from the one in \cref{Fig2}a. 
The overall influence of electroconvection is lower, with a $\Delta \Tilde{t}$ of $1.3\%$ at most. 
In contrast to the strongly nonlinear regime, for $\Tilde{\zeta}=1$, $\Delta \Tilde{t}$ hardly depends on the aspect ratio but shows a maximum over the scaled pore radius which is still present at a moderate wall potential $\Tilde{\zeta}=4$ (SI 2). 
Plotting $\Delta \Tilde{t}$ vs.\ $r_\mathrm{p}/\lambda$ results in a curve that strongly resembles a corresponding plot of the maximum in the induced axial flow velocity $\mathrm{max}(w)_{r,z,t}$, the highest velocity occurring over time anywhere in the pore (\cref{Fig3}b). This indicates that the induced velocity is a valid predictor for the effect of electroconvection. 
To understand how electroconvection affects pore charging, we analyze the flow dynamics, the flow profile, and derive an analytical model for the linear and weakly nonlinear regime. While some simplifying model assumptions break down in the strongly-nonlinear regime, the analysis provides insight into the general physical mechanism of flow induction. 

\section*{Flow induction and profile}
When the potential is applied at $t=0$, the electric field extending from the pore wall to the reservoir has a substantial axial component $E_\mathrm{z}$ where the pore meets the reservoir. 
This axial field $E_\mathrm{z}$
acts on the space charge $\rho_v$ developing near the pore wall. In turn, an axial electric body force $\rho_v E_\mathrm{z}$ drives an electroosmotic flow $w_\mathrm{E}$ into the pore.
The situation is similar to classical electroosmosis in pores, where an axial driving field is superposed on radial EDLs. 
The flow direction into the pore establishes independently of the sign of the applied potential, since both the space charge $\rho_v$ and the electric field $\mathbf{E}$ switch sign with the wall potential. 
As the electric body force drives liquid into the dead-end pore, a pressure gradient emerges that drives a compensating flow $w_\mathrm{P}$ out of the pore \cite{Levich.1962,Alizadeh.2019}. Due to the small pore diameter, the nonlinear inertial forces and transient terms in the NS equations are negligible, and the axial flow field is a superposition of the electroosmotic and pressure driven flows, $w=w_\mathrm{E}+w_\mathrm{P}$. 
Under the long-wavelength approximation, axial gradients are small compared to radial ones and the flow field can, to a good approximation, be expressed as $w(r,z,t)=w^*(r)f(z,t)$. To this end, we first derive an expression for the flow profile over the radial coordinate $w^*(r)$.

In analogy to classical electroosmosis, and because we use a simple scaling expression for the axial electric field, we assume that
$E_\mathrm{z}$ is roughly uniform in the radial direction. Then, the electroosmotic flow profile in a cylindrical pore is $w_\mathrm{E}=E_\mathrm{z}\varepsilon\left[\phi(r)-\zeta_0 \right] /\mu$ \cite{Hunter.1981}. Solutions of the PNP equations in the linear regime show that the space charge in developing EDLs largely follows its final profile scaled by a time-dependent prefactor \cite{Bazant.2004}. Building on this insight, we use the analytical solution to the Poisson-Boltzmann equation under the Debye-Hückel approximation $\zeta_0 e/(k_\mathrm{B} T)<1$ and find the electroosmotic flow profile (SI 3.B)
\begin{equation}
    w_\mathrm{E}(r)=\frac{E_\mathrm{z}\varepsilon \zeta_0}{\mu}\left[ \frac{\mathrm{I}_0(r/\lambda)}{\mathrm{I}_0(r_\mathrm{p}/\lambda)}-1\right]\text{,}
\end{equation}
where $\mathrm{I}_n(\cdot)$ is the $n$th-order modified Bessel function of the first kind. 

\begin{figure}[t]
    \centering
    \includegraphics[width=0.95\linewidth]{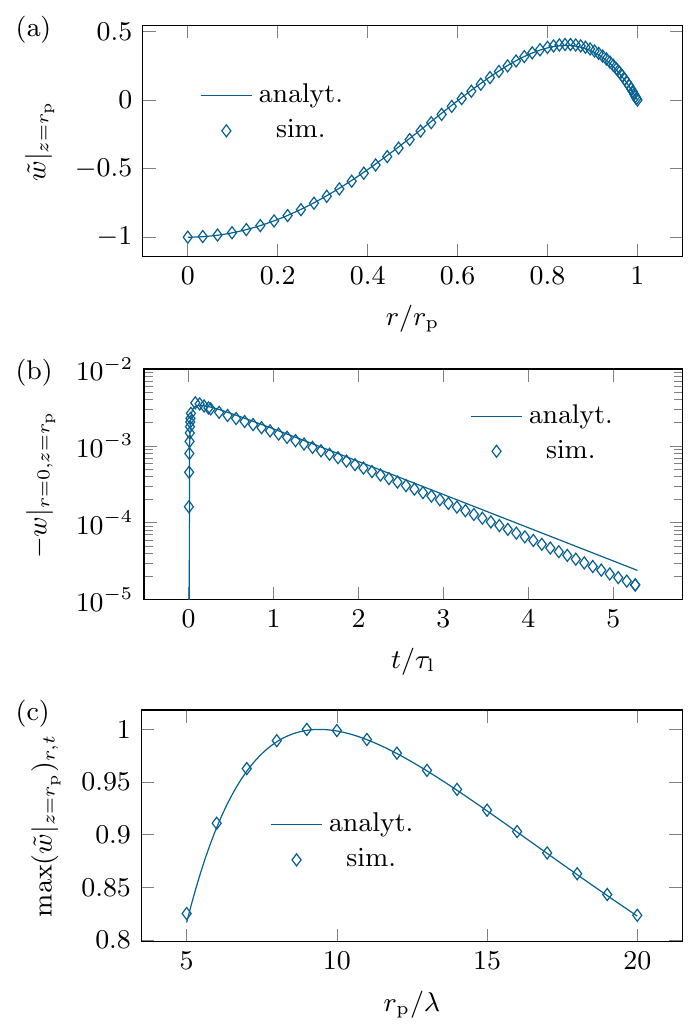}
    \caption{Comparison of the model of the flow field, \cref{eq_w_axial} (a) and \cref{eq_w(rzt)} (b, c) , (lines) and the simulation results (symbols) in the weakly nonlinear regime $\Tilde{\zeta}=1$. (a) Axial velocity $\Tilde{w}$, scaled to its maximum absolute value,
    over the scaled radial coordinate, exemplarily for $r_\mathrm{p}/\lambda=L_\mathrm{p}/r_\mathrm{p}=5, t=\tau_\mathrm{p}, z=r_\mathrm{p}$. 
    (b) Local induced axial velocity in $\si{\meter\per\second}$ over the scaled time for $r_\mathrm{p}/\lambda=5$ and $L_\mathrm{p}/r_\mathrm{p}=10$. (c) Maximum induced velocity at $z=r_\mathrm{p}$ throughout pore charging, scaled to its maximum absolute value,
    over the scaled pore radius for $L_\mathrm{p}/r_\mathrm{p}=5$.}
    \label{Fig4}
\end{figure}

The pressure-driven flow in the cylindrical pore follows a Hagen-Poiseulle profile, $w_\mathrm{P}(r)=w_\mathrm{P,max}\left[ 1-(r/r_\mathrm{p})^2\right]$ \cite{Bird.1960}. Mass conservation demands that the net volumetric flow into the pore of cross section $A_\mathrm{p}'$ vanishes, $\int_{A_\mathrm{p}'}w_\mathrm{E}+w_\mathrm{P}\mathrm{d}A=0$. With this condition we find the axial flow profile (SI 3.B)
\begin{multline}
    w^*(r)= 
    -\frac{E_\mathrm{z} \varepsilon\zeta_0}{\mu} \\
    \cdot \bigg\{ \frac{\mathrm{I}_0(r/\lambda)}{\mathrm{I}_0(r_\mathrm{p}/\lambda)}-1-4 
    \bigg[ 
    \frac{\lambda}{r_\mathrm{p}} 
    \frac{\mathrm{I}_1(r_\mathrm{p}/\lambda)}{\mathrm{I}_0(r_\mathrm{p}/\lambda)}-\frac{1}{2}
    \bigg] 
    \bigg[
    1-\left(\frac{r}{r_\mathrm{p}}\right)^2
    \bigg]
    \bigg\}\text{.}  \label{eq_w_axial}
\end{multline}
\Cref{Fig4}a and Fig. S8 demonstrate an excellent agreement between \cref{eq_w_axial} and the simulated flow profile at $\Tilde{\zeta}=1$.

\section*{Timescales}

Immediately after switching on the wall potential, no space charge has developed yet and the electric field extends from the pore wall to the grounded reservoir. The field has an axial component $E_\mathrm{z}$ next to the pore entrance.
With no space charge, the Poisson equation reduces to the Laplace equation, which does not have an inherent length scale. 
Hence, the axial electric field near the pore entrance emerges on a length scale comparable to the pore radius, $E_\mathrm{z}\propto-\zeta_0/r_\mathrm{p}$, as this is the only locally available length scale before EDL formation.

The flow is driven by the axial electric body force $\rho_v E_\mathrm{z}$, which is initially close to zero, as no significant space charge $\rho_v$ has developed at early times. 
As the EDL develops throughout the pore, the charge in the diffuse layer couples to the axial electric field, leading to an increase of the body force. At late times, when an equilibrium EDL has formed, electrostatic and pressure forces balance, and the driving force for the flow vanishes.
Thus, electroconvection is a transient phenomenon that only occurs at intermediate times. 

The pore's charging time, corresponding to EDL formation on its entire surface, generally depends on electromigration, diffusion, and convection. At moderate wall potentials, ions are mainly transported by diffusion and electromigration (cf. \cref{Fig3}a). 
In this case, the transmission line model (TLM) predicts that the pore charges exponentially on a timescale \cite{Levie.1963}
\begin{equation}
    \tau_\mathrm{TLM}=\frac{1}{2}\frac{\lambda}{r_\mathrm{p}}\frac{L_\mathrm{p}^2}{D}\text{.} \label{eq:tau_TLM}
\end{equation}
Here, we are mainly interested in the late-time response which is also influenced by the access resistance of the pore. 
Accounting for the access resistance \cite{Hall.1975} and finite pore length, the long timescale becomes \cite{Janssen.2021,Posey.1966}
\begin{equation}
    \tau_\mathrm{l}=\frac{8}{\pi^2}\frac{\lambda}{r_\mathrm{p}}\frac{L_\mathrm{p}^2}{D}\left(1+\frac{\pi^3}{16}\frac{r_\mathrm{p}}{L_\mathrm{p}}\right)\text{.} \label{eq:tau_l}
\end{equation}
This is the timescale over which EDLs develop to screen the electric field emerging from the pore wall. Overall, the axial electric field can be approximated as 
\begin{equation}
    E_\mathrm{z}=-C\frac{\zeta_0}{r_\mathrm{p}}\mathrm{e}^{-t/\tau_\mathrm{l}},\label{eq_E_z}
\end{equation}
where $C$ is an unknown dimensionless coefficient of the order of 1 that accounts for the fact that the magnitude of $E_\mathrm{z}$ is based on scaling arguments.

PNP simulations have shown that the TLM accurately describes the charging dynamics at late times \cite{Yang.2022, pedersen2023equivalent}. However, it assumes instantaneous charging of the pore entrance and thus fails to capture early times when the diffuse space charge layer of thickness $\lambda$ emerges at the pore entrance of radius $r_\mathrm{p}$. At early times, ion migration is influenced by both these length scales and the corresponding timescale is \cite{Ratschow.2022}
\begin{equation}
    \tau_\mathrm{s}=\frac{\lambda r_\mathrm{p}}{D}\text{.}
\end{equation}

The delay between the EDL formation at the pore entrance with the short timescale $\tau_\mathrm{s}$, and attaining mechanical equilibrium inside the EDL on the long timescale $\tau_\mathrm{l}$ causes electroconvection. 
Following \cite{Ratschow.2022}, the overall time evolution is proportional to $(1-e^{-t/\tau_\mathrm{s}})e^{-t/\tau_\mathrm{l}}$, explaining why velocities are higher at the short than at the long timescale (\cref{Fig1}a). 
The fundamental mechanism of the delay driving a transient flow still applies at higher wall potentials, where convection affects pore charging more. 
However, the estimation of the long timescale with \cref{eq:tau_l} loses validity in this regime. 

\section*{Spatiotemporal flow development}
Both the pressure gradient and the flow, causing viscous forces, are reactions that arise to balance the electric body force everywhere in the fluid \cite{Levich.1962}.
Thus, to understand the spatiotemporal structure of the flow field, we need to understand the development of the electric body force in $z$-direction and over time. 
At moderate wall potentials, the spatial dependence largely follows the transmission line model \cite{Levie.1963,Janssen.2021}. Overall, we model the spatiotemporal flow field as (SI 3.C)
\begin{equation}
    w(r,z,t)=w^*(r)\mathrm{erfc}\left( \frac{z}{L_\mathrm{p}}\sqrt{\frac{\tau_\mathrm{TLM}}{t}}\right)(1-e^{-t/\tau_\mathrm{s}})e^{-t/\tau_\mathrm{l}}\text{,} \label{eq_w(rzt)}
\end{equation}
with the complementary error function $\mathrm{erfc(\cdot)}=[1-\mathrm{erf(\cdot)}]$ from the analytical solution of the TLM. 
In \cref{Fig4}b and Fig. S9 (SI 3.C) we compare the induced velocity 
predicted by \cref{eq_w(rzt)} to simulations.  
The model accurately captures the dynamics and timescales (slopes). It predicts absolute values up to a constant factor, $C$ from \cref{eq_E_z}. We find that $C=1/6$ yields a good agreement across all of our numerical simulations.

\cref{eq_w(rzt)} incorporates the Debye-Hückel linearization 
through $w^*(r)$ 
and thus only applies at low and moderate wall potentials.
In this regime,
the influence of convection shows a maximum over the scaled pore radius $r_\mathrm{p}/\lambda$ (\cref{Fig3}). 
To validate our model, we compare the maximum axial velocity for $\Tilde{\zeta}=1$ at $z=r_\mathrm{p}$ over the radial coordinate, as predicted by \cref{eq_w(rzt)}, to our numerical simulations and find excellent agreement, cf. \cref{Fig4}c (both rescaled to their respective maximum value to account for the unknown constant $C$). 
Through our model we can understand why this maximum occurs. The total electric force driving the flow is $\propto E_\mathrm{z}q$, where $q$ is the instantaneous pore charge that can be estimated by the capacitance of the pore. With decreasing  
$r_\mathrm{p}/\lambda$, the capacitance and thus the total charge $q$ decrease due to EDL overlap, which decreases both the electric force and the flow. 
With increasing $r_\mathrm{p}/\lambda$, the EDL overlap decreases and the capacitance approaches that of a flat plate $2\pi r_\mathrm{p}L_\mathrm{p}\varepsilon/\lambda$, but the electric field decreases $\propto1/r_\mathrm{p}$. These two opposing trends yield a maximum in the electric body force, and thus the flow, at $r_\mathrm{p}/\lambda\approx10$.

Based on this understanding of the flow in terms of the pore capacitance, we can also formulate hypotheses for the trends in the strongly nonlinear regime (\cref{Fig2}). 
The influence of electroconvection on the pore charging time, as captured by $\Delta \Tilde{t}$, will be larger if electroconvection is stronger, or if it lasts longer.  In the strongly nonlinear regime, salt uptake dominates the late-time charging and thus the attenuation of the flow.
The leading time scale for salt uptake is $L_\mathrm{p}^2/D$ \cite{Biesheuvel.2010}. It is usually larger than $\tau_\mathrm{l}$ and keeps the flow up for longer, which can explain the increasing trend over the aspect ratio. 
At high wall potentials, the capacitance is dominated by the densely packed layer, whose thickness and capacitance increase with bulk ion concentration $c_0$ \cite{Kilic.2007}. 
An increased capacitance and thus a stronger flow with increasing bulk ion concentration is consistent with the increasing trend over $r_\mathrm{p}/\lambda\propto \sqrt{c_0}$  (SI 4).

\section*{Convective current}
\begin{figure}[t]
    \centering
    \includegraphics[width=0.95\linewidth]{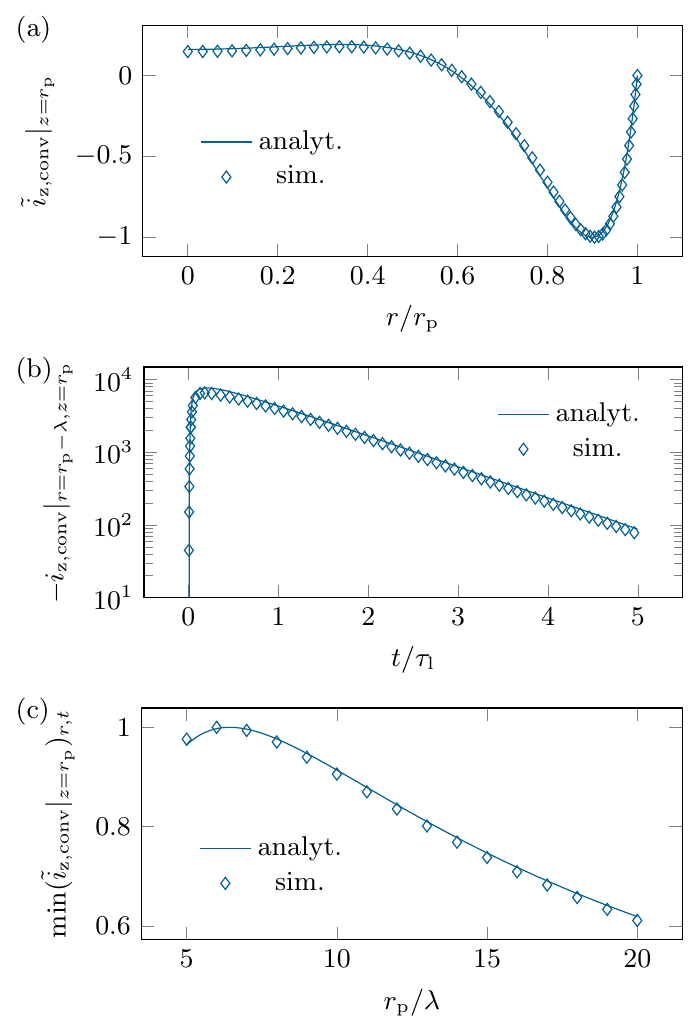}
    \caption{Comparison of the model of the convective current, \cref{eq_i_zconv}, (lines) and the simulation results (symbols) in the weakly nonlinear regime $\Tilde{\zeta}=1$. (a) Axial convective current density $\Tilde{i}_\mathrm{z,conv}$, scaled to its maximum absolute value, over the scaled radial coordinate, exemplarily for $r_\mathrm{p}/\lambda=L_\mathrm{p}/r_\mathrm{p}=5, t=\tau_\mathrm{p}, z=r_\mathrm{p}$.
    (b) Local convective current density in \si{\ampere\per\square\meter} over the scaled time for $r_\mathrm{p}/\lambda=5$ and $L_\mathrm{p}/r_\mathrm{p}=10$. (c) Minimum convective current density at $z=r_\mathrm{p}$ throughout pore charging, scaled to its maximum absolute value,
    over the scaled pore radius for $L_\mathrm{p}/r_\mathrm{p}=5$.}
	\label{Fig_current}
\end{figure}

Building on the flow field, \cref{eq_w(rzt)}, we can express the convective current density in the weakly nonlinear regime (SI 3.D)
\begin{multline}
    i_\mathrm{z,conv}=\rho_v(r,z,t) w(r,z,t)=\\- \frac{\varepsilon\zeta_0}{\lambda^2}\frac{\mathrm{I}_0(r/\lambda)}{\mathrm{I}_0(r_\mathrm{p}/\lambda)}\mathrm{erfc}\left( \frac{z}{L_\mathrm{p}}\sqrt{\frac{\tau_\mathrm{TLM}}{t}}\right)(1-e^{-t/\tau_\mathrm{s}}) w(r,z,t)\text{.} \label{eq_i_zconv}
\end{multline}
\Cref{Fig_current} shows that the analytical model for the axial current density of \cref{eq_i_zconv} accurately captures our numerical simulations, again up to a constant. The convective current mainly occurs within the EDL (\cref{Fig_current}a), where a substantial axial fluid flow coincides with a region of high space charge. 

Generally, convection, diffusion, and electromigration currents contribute to pore charging. Our simulation results of \cref{Fig2}a and \cref{Fig3}a indicate that the relative importance of electroconvection is minor at low wall potentials and increases substantially at moderate and high wall potentials. 
To explain this behavior, we introduce the electroconvection number as the ratio of electroconvection and electromigration currents $\mathrm{En}=i_\mathrm{z,conv}/i_\mathrm{z,cond}$. 
\Cref{eq_w(rzt),eq_i_zconv} show that the convective current density scales $\propto\varepsilon^2\zeta_0^3/(2\mu r_\mathrm{p}\lambda^2)$. The electromigration current density is given by $i_\mathrm{z,cond}=\sigma E_\mathrm{z}$, where the liquid conductivity under the Debye-H\"uckel approximation is $\sigma=\varepsilon D \lambda^{-2}$ \cite{Ratschow.2024}. With the axial electric field $E_\mathrm{z}\sim-\zeta_0/(2r_\mathrm{p})$, the electroconvection number is 
\begin{equation}
    \mathrm{En}=\frac{i_\mathrm{z,conv}}{i_\mathrm{z,cond}}=\frac{\varepsilon \zeta_0^2}{\mu D} \text{.} \label{eq_En}
\end{equation}
By definition, for $\mathrm{En}<1$, electroconvection is weak compared to conduction and has little influence on pore charging. For $\mathrm{En}>1$, the influence of electroconvection on the charging process can be substantial. 
Surprisingly, assuming Stokes-Einstein diffusivity, $\mathrm{En}=(3/2)(R_\mathrm{ion}/\lambda_\mathrm{B})(\zeta_0e/k_\mathrm{B}T)$, with the radius of hydrated ions $R_\mathrm{ion}$ and the Bjerrum length $\lambda_\mathrm{B}$, even though these length scales are usually considered to be too small to affect fluid flow (SI 3.E). 
Here, we find $\mathrm{En}=0.46$ for $\Tilde{\zeta}=1$, $\mathrm{En}=7.3$ for $\Tilde{\zeta}=4$ (SI 2), and $\mathrm{En}=46$ for $\Tilde{\zeta}=10$, respectively. The electroconvection number incorporates scaling relations derived in the linear regime. Our simulations show that they hold even in the regime of high potentials (Fig. S11). 
\cref{eq_En} enables a simple assessment of the importance of electroconvection. In addition, as shown in \cref{Fig2,Fig3}, the EDL thickness and the pore geometry influence the effect of electroconvection. At high applied wall potentials, the strongest effects are expected in the limit of long and narrow pores with thin EDLs. 

\section*{Implications}

Our analysis highlights that electroconvection generally 
coincides with nonlinearities in the PNP equations that occur beyond the thermal voltage $k_\mathrm{B} T/e\approx \SI{25}{\milli\volt}$. However, its influence is better measured by \cref{eq_En} which introduces a new voltage scale that indicates the onset of electroconvection in pores under charging and is an electrolyte property (further details in SI 3.E, including Refs.\cite{Bikerman.1933,Bikerman.1935,Bikerman.1940,Deryagin.1969,Lyklema.1995,Bazant.2004,Bazant.squires.2004,Squires.2004,Schnitzer.2012}). 
For aqueous electrolytes and small ions, as in our simulations, we find $\sqrt{\mu D/\varepsilon}\approx\SI{38}{mV}$. Typical values for supercapacitors, $\varepsilon\approx14 \varepsilon_0$, $\mu\approx\SI{200}{mPa s}$, and $D\approx\SI{1e-11}{m^2/s}$ \cite{Huang.2011,PietrzykThel.2024}, yield a voltage of $\approx\SI{130}{mV}$, substantially lower than their typical operating voltage $\approx\SI{1.5}{V}$. We thus hypothesize that electroconvection can play an important role in the charging of supercapacitors.

In addition to the applied potential, geometry determines the influence of electroconvection on pore charging, with the strongest effects expected for thin EDLs and slender pores (cf. \cref{Fig2,Fig3}). Typical Debye lengths in real supercapacitors are of the order of one nanometer or smaller. Average pore diameters vary widely, from $0.7-\SI{3}{nm}$ for carbide-derived carbon \cite{Simon.2008,chmiola_monolithic_2010} and activated carbon \cite{rong_hybrid_2015,deng_effect_2022} to tens of nanometers for carbon onions \cite{mcdonough_influence_2012} and reduced graphene \cite{awasthi_layer_2018}. Even for average pore diameters $<\SI{5}{nm}$, tails of pore size distributions can extend up to \SI{100}{nm} \cite{yang_graphene_2017}.
Pore lengths are usually not reported, as they are ill-defined in disordered porous media. However, for wood-derived materials, electrode layer thicknesses in excess of \SI{1}{mm} have been reported \cite{wang2021electrode}. As wood nanopores are contained inside the cellulose fibers, which can be of millimeter length, very large pore aspect ratios can be hypothesized in such structures.
Anodized metal
electrodes, as used in supercapacitors, contain highly ordered cylindrical pores with diameters of $\sim\SI{100}{nm}$ and lengths of up to hundreds of micrometers \cite{raut_vertically_2016,raj_reviewadvent_2018,deen_effect_2014,xiao_tio2_2007,ali_yahia_effect_2012}.
The fact that corresponding structures are being studied in an application context, with pore diameters of some 10 nanometers or hypothetically large aspect ratios (exceeding those considered in the present paper), suggests that convective effects can become important for the capacitive charging of realistic pores. 

In smaller pores with diameters $<\SI{1}{nm}$, the trend from our continuum-mechanical simulations suggests that the influence of convection is likely marginal. There, the pore diameter becomes comparable to the ion size, changing the physical behavior. We have not explored this limit, as it lies beyond the validity of continuum theory.
The maximum induced flow velocities in our simulations are $\approx\SI{10}{cm/s}$, exceeding those observed in electrodialysis or induced-charge electroosmosis \cite{Rubinstein.2000,Nikonenko.2017,Squires.2004}. Still, due to the small pore diameter, the Reynolds number remains small. 

Electroconvection during pore charging could be experimentally quantified through fluorescence recovery after photobleaching (FRAP) (SI 5) \cite{loren2015fluorescence,drake2024effect}. 

\section*{Conclusion}
Convection can strongly affect the capacitive charging dynamics of porous electrodes. 
The results for single pore charging from our simulations show that the change in charging time due to convection can reach values of up to $90\%$ for thin EDLs and slender pores. 
We explained the physical mechanism of delay effects leading to electroconvection during pore charging and formulated an analytical model that predicts the fluid flow up to a constant.
The model is applicable up to the weakly nonlinear regime, in which the influence of convective effects has a maximum as a function of the ratio of the pore radius and the Debye length, in agreement with the model predictions. 
For higher potentials, linearity, as assumed in the model, breaks down, but the general mechanism of flow induction still holds. 
We introduced an electroconvection number to assess the potential influence of convection on pore charging. 
Simulation results for wall potentials up to \SI{250}{mV} were obtained, much smaller than the potentials commonly applied to supercapacitors. The increasing importance of convection with increasing wall potential suggests that this mechanism, so far overlooked in the charging of supercapacitors, may be even more important in applications.

\section*{Materials and Methods}

\subsection*{Governing equations and boundary conditions}
We simulate the ion concentrations $c_{\pm}(\mathbf{r})$, ion fluxes $\mathbf{J}_\pm(\mathbf{r})$, 
electrostatic potential $\phi(\mathbf{r})$, velocity field $\mathbf{u}(\mathbf{r})$, and pressure $p(\mathbf{r})$ through the fully coupled modified Poisson-Nernst-Planck (PNP) and Navier-Stokes (NS) equations, 
\begin{subequations}\label{eq:governing_equations}
\begin{align}
	 &-\varepsilon \nabla^2\phi = e\left(c_+ - c_-\right) \text{,} \label{eq:Poisson}\\
	 &\frac{\partial c_\pm}{\partial t}  = - \nabla \cdot \mathbf{J}_\pm \text{,} \label{eq:ion_conservation}\\
	 &\mathbf{J}_\pm = -D \nabla c_\pm \mp \frac{D}{k_\mathrm{B} T}  e c_\pm \nabla\phi -  \frac{a^3 Dc_\pm \nabla\left(c_+ + c_-\right)}{1 - c_+ a^3 - c_-a^3 }  + \mathbf{u}c_\pm \label{eq_species_flux} \text{,} \\
	 &\rho_\mathrm{m} \frac{\text{D} \mathbf{u}}{\text{D}t}= -\nabla p + \mu \nabla^2 \mathbf{u} - e\left(c_+ - c_-\right) \nabla\phi \text{,} \label{eq_momentum}\\
	 &\nabla \cdot \mathbf{u} = 0 \text{,} \label{eq:incompressibility}
\end{align}
\end{subequations}
where $\text{D}/\text{D}t= \left[\partial/\partial t + \left(\mathbf{u} \cdot \nabla \right) \right] $. 
The first three lines of \cref{eq:governing_equations} were proposed by Kilic et al. \cite{Kilic.2007b} to describe the evolution of ion concentrations in a stationary background fluid. They derived the second to last term in \cref{eq_species_flux} to account for finite ion size, important in regions of high ion concentrations.
The last two lines represent the NS equations for an incompressible Newtonian liquid \cite{Probstein.1994}.
These equations are coupled through the electric body force $-e\left(c_+ - c_-\right)\nabla\phi$ in the momentum transport equation \eqref{eq_momentum} (here, $-\nabla\phi=\mathbf{E}$ is the electric field), which induces flow, and 
through the convective flux term  $\mathbf{u}c_\pm$.
Even though in some parameter regimes investigated here, the physics can be accurately described by the standard PNP and the NS equations, we use the full set of \cref{eq:governing_equations} for all simulations, as they represent the most general case we consider, and only moderately increase the computational effort in the linear regime.

We solve \cref{eq:governing_equations} on a two-dimensional, axisymmetric geometry that represents a cylindrical dead-end pore of length $L_\mathrm{p}$ and radius $r_\mathrm{p}$, and an adjacent reservoir as a quadrant of radius $r_\infty$, see Fig. S1. The radius $r_\infty$ is chosen large enough to suppress finite-size effects and resemble an infinite reservoir (SI 1.B).
The edge between pore and reservoir is blunted with a radius $r_\mathrm{blunt}=\SI{1}{nm}$. The origin of the $(r,z)$-coordinates is chosen such that $z$ measures the depth into the pore. 

To apply boundary conditions, we divide the boundary into four sections, see Fig. S1.
At the pore wall, including the end of the pore (boundary $1$), we apply a steplike wall potential $\phi=\zeta_0{\Theta}(t)$ using the Heaviside function ${\Theta}(t)$. 
Furthermore, the pore walls are considered impermeable for ions, $\mathbf{n} \cdot \mathbf{J}_\pm = 0$, and nonslipping, $\mathbf{u} = \mathbf{0}$, where $\mathbf{n}$ is the wall normal unit vector pointing to the solid.
The reservoir wall (boundary $2$) is considered uncharged, $\mathbf{n} \cdot \nabla\phi = 0$, and, like the pore walls, impermeable for ions, $\mathbf{n} \cdot \mathbf{J}_\pm = 0$, and nonslipping, $\mathbf{u} = \mathbf{0}$. 
The circular-arc reservoir boundary (boundary $3$) represents the reservoir far away from the pore, which we consider grounded, $\phi=0$, at bulk concentration, $c_\pm=c_0$, and stress-free, $\left(-p\mathbf{I}+\mu\nabla \mathbf{u}\right) = \mathbf{0}$.
At the symmetry axis (boundary $4$), we apply symmetry conditions to all equations, which corresponds to zero normal electric field, $\mathbf{n} \cdot \nabla\phi=0$, zero ion and mass flux, $\mathbf{n} \cdot \mathbf{J}_\pm = 0$ and $\mathbf{n} \cdot \mathbf{u} = 0$, and zero tangential stress, $\mathbf{t} \cdot \left(-p\mathbf{I}+\mu\nabla \mathbf{u}\right) = 0$, with the tangential vector $\mathbf{t}$.

Initially, at time $t=0$, we set the velocity, pressure and potential to zero and both cationic and anionic concentrations to $c_0$ everywhere in the computational domain. 

\section*{Data Availability}

Methods and parameters required
to reproduce this work are described in the paper and the supporting
information. The primary data underlying all presented results, the numerical
simulation models used to obtain the results (including the numerical grids),
and the code used to evaluate the analytical model are available on TUdatalib:
\url{https://doi.org/10.48328/tudatalib-2035}

\begin{acknowledgments}
We thank Timur Aslyamov for inspiring discussions and Benno Liebchen, David Fertig, Alexander Forse, and Amrita Jain for useful comments on this work. 
M.J. was supported by a Researcher Project for Young Talents grant from The Research Council of Norway (Project No. 345079). A. D. Ratschow was supported by the European Union’s Horizon 2020 Research and Innovation Program (Grant Agreement No. 883631).

A.D.R. proposed the work, A.D.R., M.J., and S.H. devised the framework, A.J.W. performed the numerical simulations, A.D.R. developed the theoretical framework, A.J.W. and A.D.R. developed the analytical model with input from M.J., all authors contributed to the interpretation of the results and prepared the manuscript, and S.H. supervised the work.
\end{acknowledgments}

\clearpage
\includepdf[pages=1]{SI}
\clearpage
\includepdf[pages=2]{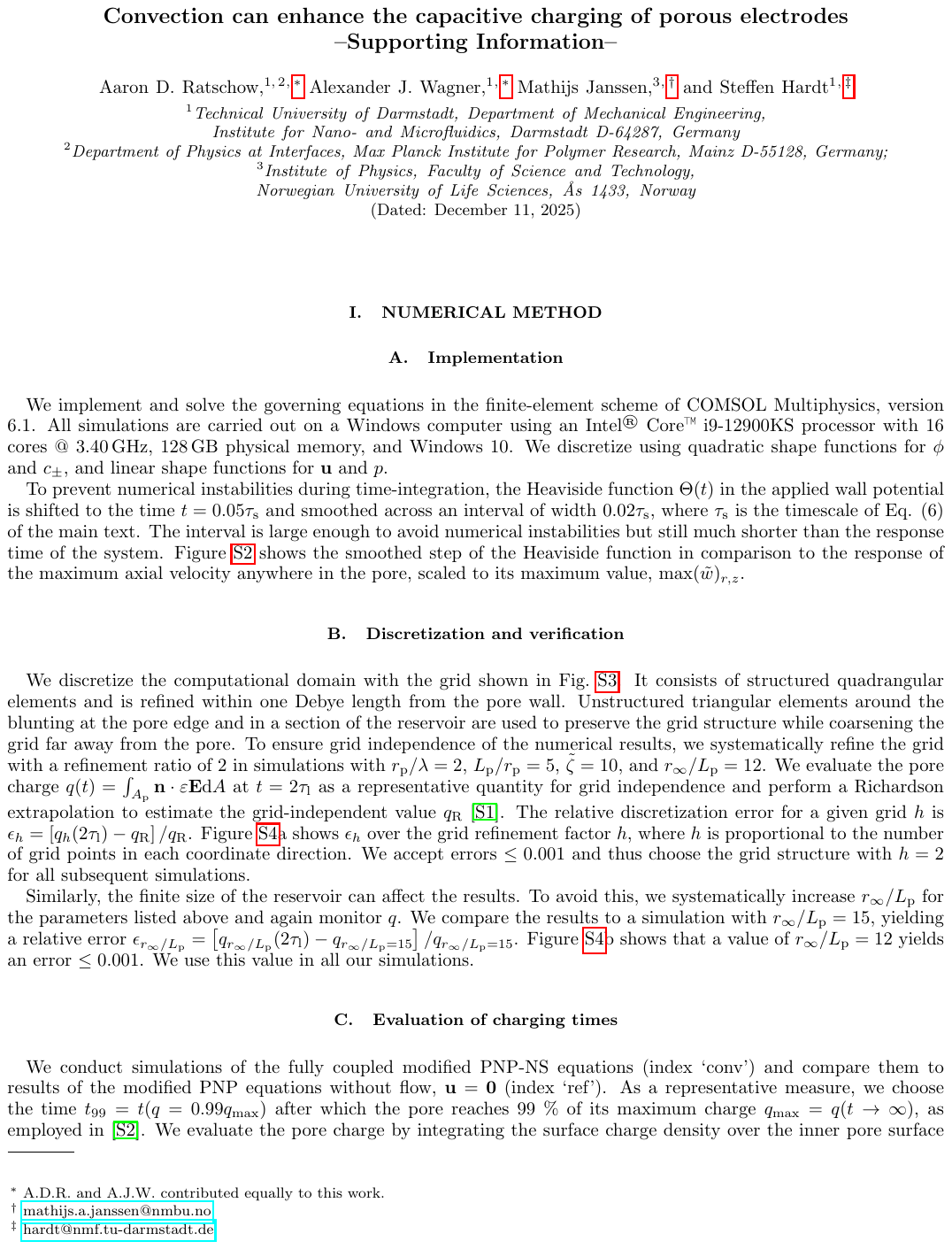}
\clearpage
\includepdf[pages=3]{SI.pdf}
\clearpage
\includepdf[pages=4]{SI.pdf}
\clearpage
\includepdf[pages=5]{SI.pdf}
\clearpage
\includepdf[pages=6]{SI.pdf}
\clearpage
\includepdf[pages=7]{SI.pdf}
\clearpage
\includepdf[pages=8]{SI.pdf}
\clearpage
\includepdf[pages=9]{SI.pdf}
\clearpage
\includepdf[pages=10]{SI.pdf}
\clearpage
\includepdf[pages=11]{SI.pdf}
\clearpage
\includepdf[pages=12]{SI.pdf}
\clearpage
\includepdf[pages=13]{SI.pdf}
\clearpage
\includepdf[pages=14]{SI.pdf}
\clearpage
\includepdf[pages=15]{SI.pdf}
\clearpage
\includepdf[pages=16]{SI.pdf}
\clearpage
\includepdf[pages=17]{SI.pdf}
\clearpage
\includepdf[pages=18]{SI.pdf}
\clearpage
\includepdf[pages=19]{SI.pdf}
\clearpage
\includepdf[pages=20]{SI.pdf}
\clearpage
\includepdf[pages=21]{SI.pdf}

\end{document}